\documentclass[letterpaper, 10 pt, conference]{ieeeconf}  
\usepackage{mathtools}
\usepackage{amsmath}
\usepackage{color}
\usepackage{cuted}
\usepackage{multirow}
\usepackage{booktabs}
\usepackage{epstopdf}

\IEEEoverridecommandlockouts                              

\title{\LARGE \bf
Robust Estimation of Battery State of Health Using Reference Voltage Trajectory
}

\author{Rui Huang$^{1}$, Jackson Fogelquist$^{1}$, and Xinfan Lin$^{1*}$ 
\thanks{$^{1}$R. Huang, J. Fogelquist and X. Lin are with the Department of Mechanical and Aerospace Engineering, University of California, Davis, CA 95616, USA, corresponding author e-mail: 
        {\tt\small lxflin@ucdavis.edu}}%
}

\begin{document}

\maketitle
\thispagestyle{empty}
\pagestyle{empty}

\begin{abstract}

Accurate estimation of state of health (SOH) is critical for battery applications.
Current model-based SOH estimation methods typically rely on low C-rate constant current tests to extract health parameters like solid phase volume fraction and lithium-ion stoichiometry, which are often impractical in real-world scenarios due to time and operational constraints.
Additionally, these methods are susceptible to modeling uncertainties that can significantly degrade the estimation accuracy, especially when jointly estimating multiple parameters.
In this paper, we present a novel reference voltage-based method for robust battery SOH estimation.
This method utilizes the voltage response of a battery under a predefined current excitation at the beginning of life (BOL) as a reference to compensate for modeling uncertainty.
As the battery degrades, the same excitation is applied to generate the voltage response, which is compared with the BOL trajectory to estimate the key health parameters accurately.
The current excitation is optimally designed using the Particle Swarm Optimization algorithm to maximize the information content of the target parameters.
Simulation results demonstrate that our proposed method significantly improves parameter estimation accuracy under different degradation levels, compared to conventional methods relying only on direct voltage measurements.
Furthermore, our method jointly estimates four key SOH parameters in only 10 minutes, making it practical for real-world battery health diagnostics, e.g., fast testing to enable  battery repurposing.
\end{abstract}

\section{INTRODUCTION}
\label{sec:introduction}

  
 


Accurate monitoring of the state of health (SOH) of lithium-ion batteries is essential for ensuring their safety, reliability, and efficacy in applications, including electric vehicles (EVs) and energy storage systems \cite{xiong2018towards}.
Recently, there is an emerging demand for fast and accurate estimation of battery SOH to enable battery repurposing at a large scale \cite{neubauer2015identifying}.    
Model-based SOH estimation \cite{lin2019modeling} has been widely used due to its ability of estimating internal battery parameters directly indicating SOH.
This practice involves finding the model parameters that best fit the model output with the measurement directly.
The results, however, are heavily affected by the inevitable modeling uncertainties, i.e., the mismatch between the true dynamics of the battery and those captured by the model, even if the model is correctly parameterized \cite{fogelquist2023error}.

It is noted that SOH indicators are typically associated with the parameters of the battery open-circuit voltage (OCV), such as the solid-phase active material volume fraction $\varepsilon_s$ and lithium-ion stoichiometry $\beta$, which are directly linked to the battery's capacity and ability to store and deliver energy. 
For example, in \cite{lee2020electrode}, the electrode capacity and stoichiometry were estimated as SOH indicators by fitting the model output to low C-rate data (1/20 C).
Schmitt et al. \cite{schmitt2023capacity} proposed to use 1/15 C current to reconstruct OCV and estimate capacity based on charging/discharging data at rates lower than 1/4 C.
Rahman et al. \cite{rahman2022li} explored the use of Constant-Current-Constant-Voltage (CC-CV) charging/discharging protocol (0.7 C) to estimate battery electrode SOH,  
but with additional instrumentation such as the case reference electrodes. 
It is seen that these methods typically rely on low C-rate constant current testing over a wide range of battery state-of-charge (SOC) 
to measure the battery OCV with good fidelity, resulting in an extended testing duration.
However, such slow tests are rarely feasible in practical applications where timely assessment within a short testing period is necessary,
and hence higher C-rate data need to be used.
Under such conditions, the voltage response of the battery is significantly influenced by dynamic effects such as polarization and diffusion 
\cite{qiu2019polarization}, and deviates from the open circuit condition.
It should be noted that these dynamic effects are very difficult to be modeled exactly, and the residual, which we refer to as the model uncertainty 
will 
make it extremely challenging to accurately retrieve the OCV-related parameters.


In this paper, we address these challenges by proposing a reference voltage-based method that can compensate for modeling uncertainty.
The key idea is to measure the voltage response of a battery under a designed current excitation at the beginning of life (BOL) to establish a reference voltage trajectory.
This reference, combined with the voltage response of BOL battery model under the same input, 
inherently captures the modeling uncertainties 
under the test setup.
As the battery degrades, the same excitation is applied to generate the voltage response, which is compared with the BOL trajectory to estimate the key health parameters accurately.
To further enhance parameter identifiability and estimation accuracy, we design optimal current excitation profiles using Particle Swarm Optimization (PSO), which maximize the information content of the target parameters in the voltage responses.
Simulation studies conducted under various degradation levels demonstrate that our proposed reference voltage method significantly improves estimation accuracy compared to conventional approaches that estimate directly from voltage measurements.
Our approach enables the estimation of four key battery health parameters within a short testing time of 10 minutes.
The proposed method's enhanced accuracy, speed and ease of implementation make it particularly advantageous for practical applications, e.g., the repurposing of batteries for second-life use, where quick and accurate SOH assessment and testing are crucial.

\section{BATTERY MODELING}


We use the single particle model with electrolyte dynamics (SPMe) as the model for battery SOH estimation as it strikes a good balance between accuracy and computational cost \cite{moura2016battery}.
The model is a simplification of the full-order first-principle Doyle-Fuller-Newman (DFN) model, where the diffusion of lithium-ion in all solid particles across an electrode is represented by one single spherical particle, neglecting the spatial difference between different particles and assuming uniform reaction current density along the electrode thickness direction.
In SPMe, the input is the current $I$, and the output voltage $V$ is mainly comprised of four parts, as shown in \eqref{eq:SPMe output},
where the subscript $_n$ denotes the variables of the negative electrode (anode) and $_p$ denotes those of the positive electrode (cathode). 
The first one is the difference in open circuit potential (OCP) $U$ between the two electrodes, which are nonlinear functions of the solid particle surface lithium-ion concentration $c_{se}$ driven by solid phase diffusion.
The second one is the difference between electrodes in overpotential $\eta$, which drives the intercalation/deintercalation of lithium at the particle surface through the Butler-Volmer equation.
The third one $\phi_e$ is the electrolyte potential (difference) due to the concentration gradient of lithium-ion between the two electrodes, driven by electrolyte lithium diffusion.
Finally, the term $IR_l$ represents the ohmic effect of the lumped battery resistance, including the contact resistance, electrolyte resistance and that of the solid-electrolyte interphase (SEI).
\begin{equation}
\label{eq:SPMe output}
\begin{split}
    &V = \\ &(U_p(c_{se,p}) - U_n(c_{se,n})) + (\eta_p(c_{se,p}) - \eta_n(c_{se,n}))  \\
&+ (\phi_{e,p}(c_{e,p}) - \phi_{e,n}(c_{e,n})) - IR_l
\end{split}
\end{equation}

It is noted that under the conventional low C-rate testing schedule, the battery is close to equilibrium, that is, the overpotential $\eta$, electrolyte potential $\phi_e$ and ohmic voltage drop $IR_l$ are all negligible, and the concentration of lithium-ion in the particle is uniform.
As a result, the output voltage is dominated by OCP, resulting in the battery OCP model, as shown in \eqref{eq:OCP output}. 
\begin{equation}
\label{eq:OCP output}
    V \approx U_p(c_{se,p}) - U_n(c_{se,n}).
\end{equation}
Details of the dynamic voltage components can be found in literature \cite{moura2016battery, lai2020analytical}, and are omitted here. 

In \eqref{eq:OCP output}, $c_{se}$ is the same as the average lithium ion concentration in the solid particle $c_s$, which can be computed by the battery SOC using the Coulomb counting method
\begin{equation}
\label{eq:theta dynamics}
    {\rm{SOC}} = {\rm{SOC}}_0 - \frac{\int_0^t I(\tau)d\tau}{F A \delta c_{s,max} \varepsilon_{s} (\beta_{100\%}-\beta_{0\%})},
\end{equation}
where ${\rm{SOC}}_0$ is the initial SOC, $F$ stands for the Faraday constant,
$A$ and $\delta$ are electrode area and thickness respectively, $c_{s,max}$ stands for the maximum lithium concentration of the electrode and $\varepsilon_s$ denotes the active material volume fraction.
In addition, we use $\beta_{0\%}$ and $\beta_{100\%}$ to represent the  stoichiometry limit at 0\% and 100\% SOC. 
The $c_{se}$ can be computed based on \eqref{eq:beta_0}:
\begin{equation}
\label{eq:beta_0}
    c_{se} = [\beta_{0\%} + (\beta_{100\%} - \beta_{0\%}) \cdot {\rm{SOC}}] \cdot c_{s,max}.
\end{equation}

The target parameters for estimation studied in this paper include the active material volume fraction and lithium-ion stoichiometry at both positive and negative electrodes when the battery is fully discharged, namely $\varepsilon_{s,n}$, $\varepsilon_{s,p}$, $\beta_{n,0\%}$ and $\beta_{p,0\%}$.
Among them, $\varepsilon_{s}$ directly quantifies the amount of active material in the electrode.
As batteries age, active material will decrease due to mechanical stresses, particle cracking and/or dissolution \cite{o2022lithium}.
A decrease in $\varepsilon_{s}$ hence directly reflects the loss of active material (LAM), which is a key battery physical degradation mechanism leading to reduced capacity and power.
Meanwhile, the stoichiometry ratios define the battery electrochemistry operation range.
Specifically, $\beta_{0\%}$ and $\beta_{100\%}$ represent the proportion of lithium in the electrode lattice structure 
when the battery OCV is at the lower and upper limit of the voltage operating range. 
As battery ages, the stoichiometry ratios at the two ends will shift, caused by loss of lithium inventory (LLI) due to side reactions such as SEI formation and lithium plating, 
causing capacity loss \cite{han2019review}.
Therefore, estimating the change of these parameters enables the identification of both LAM and LLI, the two primary degradation mechanisms in lithium-ion batteries,  \cite{birkl2017degradation} and hence the monitoring of battery SOH.

Conventionally, the estimation of these OCV parameters is performed using low C-rate test data, i.e., fitting input $I$ and output $V$ using \eqref{eq:OCP output}, \eqref{eq:theta dynamics} and \eqref{eq:beta_0},
as the voltage output is closer to OCV under smaller current.
The tests also need to cover a wide range of SOC (which can be defined by $\beta$) for better identifiability of the capacity-related $\varepsilon_s$ \cite{lin2015analytic, lee2019estimation}, leading to long testing time.
Attempts to reduce testing time by increasing the current would inevitably deviate from the equilibrium condition and 
introduce the aforementioned battery dynamics (polarization, diffusion, etc.) into the voltage measurement,
necessitating the use of dynamic battery models for estimation. 
Nevertheless, no model can capture the true battery dynamics perfectly. 
This is especially true for the control-oriented models, such as the equivalent circuit model and SPMe, 
which involves model reduction techniques to reduce the model complexity. 
The resultant modeling uncertainties can significantly undermine the estimation accuracy of these parameters, especially in the joint estimation scenario involving multiple parameters \cite{fogelquist2023error}.
This challenge has inspired us to develop a methodology that can 
address the limitations of conventional test methods.

\section{PROPOSED METHODOLOGY} 

%
 
\begin{figure}[thpb]
      \centering
      \includegraphics[scale=0.7]{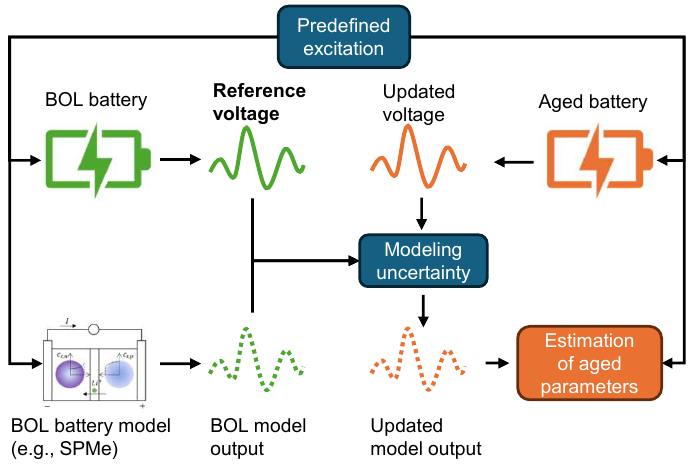}
      \caption{Schematic of reference voltage-based SOH estimation method}
      \label{fig:schematic}
\end{figure}

As presented in previous sections, the modeling uncertainty presents a major challenge to the fast and accurate estimation of $\varepsilon_s$ and $\beta_{0\%}$.
In this research, we proposed a simple but effective reference voltage-based method to address this issue.
Compared with the existing practice, the method only requires an additional step of applying a predefined current excitation to the battery at the beginning of life to generate a reference voltage trajectory. 
Meanwhile, the voltage prediction of the battery model at BOL under the nominal parameter set 
is also recorded.
The difference between the measured and predicted voltage provides an evaluation of the modeling uncertainty.
When the battery degrades and needs re-evaluation of SOH, we will apply the same excitation to the battery to obtain the updated response.
The parameters can be estimated using the new voltage data while the previous reference compensates for the model uncertainty.
The schematics of the proposed method is shown in Fig. \ref{fig:schematic}, with details discussed as follows.

\subsection{Formulation of Reference Voltage-based SOH Estimation}

Here, we provide an explanation of the mechanism by which the reference voltage-based method can significantly reduce the estimation error.
The measurement of a battery output voltage trajectory over time $\boldsymbol{V}^m$ can be expressed as
\begin{equation}
    \boldsymbol{V}^m (\boldsymbol{\theta}, \boldsymbol{x_0}, \boldsymbol{u}, t) = 
    \boldsymbol{S}(\boldsymbol{\theta}, \boldsymbol{x_0}, \boldsymbol{u}, t) + \boldsymbol{\delta}^{model}(t) + \boldsymbol{\delta}^{meas}(t),
\end{equation}
where $\boldsymbol{S}(\boldsymbol{\theta}, \boldsymbol{x_0}, \boldsymbol{u}, t)$ is the output prediction of a model (e.g., SPMe), 
depending on the model parameters $\boldsymbol{\theta}$, initial states of the model $\boldsymbol{x_0}$ and input current $\boldsymbol{u}$.
In addition, $\boldsymbol{\delta}^{model}(t)$ represents the uncertainty between the adopted model and the true system dynamics, 
and $\boldsymbol{\delta}^{meas}(t)$ denotes the measurement uncertainty, including the sensor drift and noise. 
 
At the BOL of the battery, 
the output measurement $\boldsymbol{V}_0^m$ is parameterized by the nominal battery parameters $\boldsymbol{\theta}_0$ as 
\begin{equation}
\label{eq:V_0_m}
    \boldsymbol{V}_0^m (\boldsymbol{\theta}_0, \boldsymbol{x_0}, \boldsymbol{u},t) = 
    \boldsymbol{S}(\boldsymbol{\theta}_0, \boldsymbol{x_0}, \boldsymbol{u},t) + 
    \boldsymbol{\delta}_0^{model}(t) + \boldsymbol{\delta}_0^{meas}(t).
\end{equation}
When the battery degrades and its parameters change to $\boldsymbol{\theta}_1$, the output measurement $\boldsymbol{V}_1^m$ under the same input $\boldsymbol{u}$ changes to
\begin{equation}
\label{eq:V_1_m}
    \boldsymbol{V}_1^m (\boldsymbol{\theta}_1, \boldsymbol{x_0}, \boldsymbol{u},t) = 
    \boldsymbol{S}(\boldsymbol{\theta}_1,\boldsymbol{x_0}, \boldsymbol{u},t) + 
    \boldsymbol{\delta}_1^{model}(t) + \boldsymbol{\delta}_0^{meas}(t).
\end{equation}

The conventional practice of estimating the aged battery parameter $\boldsymbol{\theta}_1$ is performed using only the measurement of $\boldsymbol{V}_1^m$ at discrete time instants $k=1,2,...N$, for example, in the classical least squares form as
\begin{equation}
\label{eq:conventional opt}
\begin{split}
    \min_{ \boldsymbol{\hat \theta}_1} \sum_{k=1}^{N} \left[
    \boldsymbol{V}_{1}^m(\boldsymbol{\theta}_1, \boldsymbol{x_0}, \boldsymbol{u}, t_k) - 
    \boldsymbol{S}(\boldsymbol{\hat \theta}_1, \boldsymbol{x_0}, \boldsymbol{u}, t_k) \right]^2 \\
    = \sum_{k=1}^{N} \bigg[ \boldsymbol{S}(\boldsymbol{\theta}_1,\boldsymbol{x_0}, \boldsymbol{u},t_k) 
        - \boldsymbol{S}(\boldsymbol{\hat \theta}_1, \boldsymbol{x_0}, \boldsymbol{u}, t_k) \\
       + \boldsymbol{\delta}_1^{model}(t_k) + \boldsymbol{\delta}_1^{meas}(t_k) \bigg]^2 
\end{split}
\end{equation}
It is seen that if the model uncertainty $\boldsymbol{\delta}_1^{model}$ and measurement uncertainty $\boldsymbol{\delta}_1^{meas}$ are zero, the least squares estimate will yield the correct results $\boldsymbol{\hat \theta}_1 = \boldsymbol{\theta}_1$,
which minimize the square errors to 0. 
However, in practice when the presence of these uncertainties is inevitable, the estimation results will deviate from the 
actual value \cite{fogelquist2023error}.


We assume that the $\boldsymbol{\delta}^{model}$ is the same for models parameterized by different $\boldsymbol{\theta}$, while $\boldsymbol{\delta}^{meas}$ differs in different measurements.   

For the proposed reference voltage-based method, suppose we have obtained the voltage measurement at the BOL $\boldsymbol{V}_0^m$ beforehand (e.g., from the battery manufacturer or through benchmark testing), and we aim to estimate the new parameter $\boldsymbol{\hat \theta}_1$ using both $\boldsymbol{V}_0^m$ and $\boldsymbol{V}_1^m$, and knowledge of (nominal) parameters $\boldsymbol{\theta}_0$.
By subtracting \eqref{eq:V_0_m} from \eqref{eq:V_1_m}, 
the difference between $\boldsymbol{V}_1^m$ and $\boldsymbol{V}_0^m$ takes the form
\begin{equation}
    \begin{split}
        \boldsymbol{V}_1^m &(\boldsymbol{\theta}_1, \boldsymbol{x_0}, \boldsymbol{u},t) -  
        \boldsymbol{V}_0^m (\boldsymbol{\theta}_0, \boldsymbol{x_0}, \boldsymbol{u},t)  \\
        =& \boldsymbol{S}(\boldsymbol{\theta}_1,\boldsymbol{x_0}, \boldsymbol{u},t) - 
        \boldsymbol{S}(\boldsymbol{\theta}_0,\boldsymbol{x_0}, \boldsymbol{u},t) \\
        +& \boldsymbol{\delta}_1^{model}(t) -  \boldsymbol{\delta}_0^{model}(t)
	+ \boldsymbol{\delta}_1^{meas}(t) - \boldsymbol{\delta}_0^{meas}(t),
    \end{split}
		\label{eq:dV}
\end{equation}

There are multiple benefits of working with the voltage difference form in (\ref{eq:dV}) for estimation.
First, the model uncertainty terms $\boldsymbol{\delta}^{model}(t)$ will be subtracted from each other and largely canceled out.
Note that the two voltage trajectories are generated under the same input current excitation, and hence $\boldsymbol{\delta}_0^{model}$ and $\boldsymbol{\delta}_1^{model}$ are expected to be fairly similar.
Though they may still depend on $\boldsymbol{\theta}$, but the impact will be much less significant than that in (\ref{eq:conventional opt}), 
where the model uncertainty $\boldsymbol{\delta}_1^{model}$ appears fully.  
Second, measurement bias, which is a common type of measurement uncertainty, will also be significantly reduced.
Intuitively, bias is the constant component of $\boldsymbol{\delta}^{meas}$, which will be completely canceled out in (\ref{eq:dV}) if the same measurement technique and equipment are used for testing.
However, it should be pointed out that the measurement variance, which is the variational part of the measurement uncertainty, can be amplified by performing the voltage subtraction.
A detailed analysis of the estimation errors will be provided in the next subsection. 

By using the voltage trajectory difference in (\ref{eq:dV}) to estimate the target parameters $\boldsymbol{\theta}$, 
the new estimation problem is reformulated as
\begin{equation}
\label{eq:new opt}
\begin{split}
    \min_{\boldsymbol{\hat \theta}_1} \sum_{k=1}^{N} 
    [ (V_{1}^m(\boldsymbol{\theta}_1, \boldsymbol{x_0}, \boldsymbol{u},t_k) - 
    V_{0}^m(\boldsymbol{\theta}_0, \boldsymbol{x_0}, \boldsymbol{u}, t_k)) - \\
    ( S(\boldsymbol{\hat \theta}_1, \boldsymbol{x_0}, \boldsymbol{u}, t_k) - 
    S(\boldsymbol{\theta}_0, \boldsymbol{x_0}, \boldsymbol{u}, t_k) ) ]^2.
\end{split}
\end{equation}



\subsection{Quantification of Estimation Error}

In this subsection, we will provide a detailed error analysis of the proposed method. 
The estimation error of battery health parameters for the reference voltage-based method can be derived as follows.
By substituting \eqref{eq:V_0_m} and \eqref{eq:V_1_m} into \eqref{eq:new opt}, the objective function becomes
\begin{equation}
\label{eq:new opt 3}
\begin{split}
    &\min_{\boldsymbol{\hat \theta}_1} J = \sum_{k=1}^{N} 
     [ S(\boldsymbol{ \theta}_1, \boldsymbol{x_0}, \boldsymbol{u}, t_k) - 
     S(\boldsymbol{\hat \theta}_1, \boldsymbol{x_0}, \boldsymbol{u}, t_k) +\\ 
		& \boldsymbol{\delta}_1^{model}(t_k) - \boldsymbol{\delta}_0^{model}(t_k) 
    + \boldsymbol{\delta}_1^{meas}(t_k) - \boldsymbol{\delta}_0^{meas}(t_k)  ]^2
\end{split}
\end{equation}

Now we lump all the uncertainty-related terms into a single term  
$\Delta \boldsymbol{\delta}$, 
before decompose it into a constant and a varying component,
\begin{equation}
\label{eq:DV1}
\begin{split}
   &\Delta \boldsymbol{\delta}(t_k) = \\&\boldsymbol{\delta}_1^{model}(t_k) - \boldsymbol{\delta}_0^{model}(t_k) + \boldsymbol{\delta}_1^{meas}(t_k) - \boldsymbol{\delta}_0^{meas}(t_k) \\
		&= \Delta \bar{\delta} + d \boldsymbol{\delta}(t_k).
\end{split}
\end{equation}
The constant component $\Delta \bar{\delta}$ captures the invariant part of the model and measurement uncertainties 
that are not completely canceled.
The varying component $d\boldsymbol{\delta}(t_k)$, on the other hand, 
mainly includes the random noises in measurement, whose distribution is determined by the accuracy of the sensor.
Suppose the random sensor noise can be described by a zero-mean i.i.d. Gaussian noise with variance $\sigma^2$, i.e. $\sim N(0,\sigma^2)$,
$\boldsymbol{\Delta} \delta(t_k)$ can then be approximated by,
\begin{equation}
\boldsymbol{\Delta \delta}(t_k) \sim N(\Delta \bar{\delta}, 2\sigma^2).
\label{eq:DVG}
\end{equation}
Note that the variance is amplified by a factor of $2$ due to the subtraction of voltage trajectories. 

Then, by applying the optimality condition $\frac{\partial J}{\partial \boldsymbol{\hat \theta}_1} = 0 $ and the first-order Taylor expansion of $S$ 
evaluated at $\boldsymbol{\hat \theta}_1$, the estimation error $\Delta \boldsymbol{\hat \theta}_1 = \boldsymbol{\theta}_1 - \boldsymbol{\hat \theta}_1$ can be derived as,
\begin{equation}
\label{eq:dtheta}
\begin{split}
    \Delta \boldsymbol{\hat \theta}_1 = -
    \left[ \sum_{k=1}^{N}
    \frac{\partial S^T}{\partial \boldsymbol{\hat \theta}_1}(t_k) 
    \frac{\partial S}{\partial \boldsymbol{\hat \theta}_1}(t_k)
    \right]^{-1}
    \sum_{k=1}^{N} 
    \frac{\partial S^T}{\partial \boldsymbol{\hat \theta}_1}(t_k) \Delta \boldsymbol{\delta}(t_k).
\end{split}
\end{equation}
Detailed derivation procedures are skipped here, which are similar to those in \cite{fogelquist2023error}. 
In (\ref{eq:dtheta}), $\frac{\partial \boldsymbol{S}}{\partial \boldsymbol{\hat \theta}_1}(t_k)$ 
represents  
the sensitivity of the model output at time $t_k$, evaluated at $\boldsymbol{\hat \theta}_1$, 
which can be computed using the physical model, with details provided in \cite{lai2020analytical}. 

Next, by plugging the statistics of $\Delta \boldsymbol{\delta}(t_k)$ in (\ref{eq:DVG}) into (\ref{eq:dtheta}), 
the mean and variance of the estimation error ${\rm E} \Delta \boldsymbol{\hat \theta}_1$ can be obtained as 
\begin{equation}
\label{eq:EV_err}
\begin{aligned}
    &{\rm E} \Delta \boldsymbol{\hat \theta}_1 = -
    \left[ \sum_{k=1}^{N}
    \frac{\partial S^T}{\partial \boldsymbol{\hat \theta}_1}(t_k) 
    \frac{\partial S}{\partial \boldsymbol{\hat \theta}_1}(t_k)
    \right]^{-1}
    \sum_{k=1}^{N} 
    \frac{\partial S^T}{\partial \boldsymbol{\hat \theta}_1}(t_k) \Delta \bar{\delta},   \\
		&\boldsymbol{cov}(\Delta \boldsymbol{\hat \theta}_1)	= 2\left[ \sum_{k=1}^{N}
    \frac{\partial S^T}{\partial \boldsymbol{\hat \theta}_1}(t_k) 
    \frac{\partial S}{\partial \boldsymbol{\hat \theta}_1}(t_k) 
    \right]^{-1}\sigma^2. 
\end{aligned}    
\end{equation}

The derived error formula confirms the previous intuitions about the proposed reference voltage-based method. 
First, the method could significantly reduce the mean (bias) of the estimation error.
It is seen from (\ref{eq:EV_err}) that ${\rm E} \Delta \boldsymbol{\hat \theta}_1$ is proportional to $\Delta \bar{\delta}$,
which is negligible compared to the original uncertainty $\boldsymbol{\delta}_1$ as it is the remainder after canceling out $\boldsymbol{\delta}_0$.
Second, the variance of the estimation error is indeed amplified due to voltage subtraction as seen by the factor $2$ in 
$\boldsymbol{cov}(\Delta \boldsymbol{\hat \theta}_1)$.
Third, the estimation error is also heavily affected by the reference voltage trajectory.
Specifically, as seen in (\ref{eq:EV_err}), both estimation bias and variance are related 
to the inverse of 
\begin{equation}
\boldsymbol{F}_{info} = \left[ \sum_{k=1}^{N}
    \frac{\partial S^T}{\partial \boldsymbol{\hat \theta}_1}(t_k) 
    \frac{\partial S}{\partial \boldsymbol{\hat \theta}_1}(t_k)
    \right],
\label{eq:FI}
\end{equation}
which is the Fisher information (matrix) of the voltage data about the target parameters 
$\boldsymbol{\hat \theta}$ \cite{fogelquist2023error}.
It is possible to design a reference voltage trajectory with maximal Fisher information
by optimizing the input excitation, to further reduce or minimize estimation errors \cite{lai2021new, huang2023reinforcement}. 
This will be especially beneficial for reducing the estimation variance, which is amplified by voltage subtraction.

		

\subsection{Design of Reference Voltage Trajectory}

In this subsection, we will explore how to design the reference voltage trajectory to optimize the estimation accuracy of the proposed method.

Based on the discussion in the previous subsection, 
an optimization problem can be formulated to maximize a specific metric of the Fisher information matrix,
\begin{equation}
\max_{\boldsymbol I(t)}J\left( \sum_{k=1}^{N}
    \frac{\partial S^T}{\partial \boldsymbol{\hat \theta}_1}(t_k) 
    \frac{\partial S}{\partial \boldsymbol{\hat \theta}_1}(t_k)
    \right).
\label{eq:OED}
\end{equation}
The optimization variables are the current excitation $\boldsymbol I(t)$, 
which will drive the sensitivity $\frac{\partial S^T}{\partial \boldsymbol{\hat \theta}_1}$ 
and hence influence the Fisher information as shown in our previous works \cite{lai2021new}. 
The metric $J$ can be selected with various options, including 
\begin{itemize}
    \item the determinant of the Fisher information matrix (D-optimality) 
    \item the minimum eigenvalue (E-optimality), and
    \item the trace of the inverse Fisher information matrix (A-optimality).
\end{itemize}

To compute the sensitivity which is needed for evaluating the Fisher information, we utilize the OCP model in \eqref{eq:OCP output}, \eqref{eq:theta dynamics} and  \eqref{eq:beta_0}.
This simple model focuses only on the components related to the target parameters, and simplifies the sensitivity calculations significantly without major losses in accuracy. 
The sensitivities of the output voltage $V$ with respect to the target parameters, namely $\varepsilon_s$ and $\beta_{0\%}$, are calculated using the chain rule as
\begin{equation}
\label{eq:sens_beta_p}
    \frac{\partial V}{\partial \beta_{p,0\%}} =  \frac{\partial U}{\partial c_{se,p}} \cdot \frac{\partial c_{se,p}}{\partial \beta_{p,0\%}} = \frac{\partial U}{\partial c_{se,p}} \cdot c_{s,p,max} \cdot (1-\rm{SOC}_0), 
\end{equation}
\begin{equation}
\label{eq:sens_beta_n}
    \frac{\partial V}{\partial \beta_{n,0\%}} = - \frac{\partial U}{\partial c_{se,n}} \cdot \frac{\partial c_{se,n}}{\partial \beta_{n,0\%}} = - \frac{\partial U}{\partial c_{se,n}} \cdot c_{s,n,max} \cdot (1-\rm{SOC}_0), 
\end{equation}
\begin{equation}
\label{eq:sens_epsilon}
    \frac{\partial V}{\partial \varepsilon_{s,i}} = -\frac{\partial U}{\partial c_{se,i}} \cdot \frac{\partial c_{se,i}}{\partial \varepsilon_{s,i}} = - \frac{\partial U}{\partial c_{se,i}} \cdot 
    \left ( \frac{\int_0^t I(\tau)d\tau}{F A_i \delta_i c_{s,i,max} \varepsilon_{s,i}^2}
    \right).
\end{equation}
where $\frac{\partial U}{\partial c_{se}}$ is the slope of the OCP curve.

The formulated Fisher information objective is optimized using Particle Swarm Optimization (PSO), 
which is a population-based meta-heuristic algorithm suitable for complex, nonlinear optimization problems.
It solves the optimization problem by simulating the social behavior of particles moving through a search space, where each particle adjusts its position and velocity based on its own experience and that of the neighboring particles.
A more detailed introduction about the implementation of PSO can be found in \cite{kennedy2006swarm}.
In this work, PSO is used to optimize the current excitation over a 600-second horizon, divided into 10 intervals of 60 seconds each, i.e., with 10 decision variables $[I_1,...,I_{10}]$ corresponding to the current values applied during each interval.
Safety constraints are imposed by bounding the current, voltage and SOC within their respective limits, ensuring safe and practical operation throughout the test.
The maximum input current C-rate is set to 1.
The initial SOC is set to 0 to standardize the starting point for the optimization.

\section{SIMULATION VALIDATION RESULTS}

\begin{figure}[thpb]
      \centering
      \includegraphics[scale=0.55]{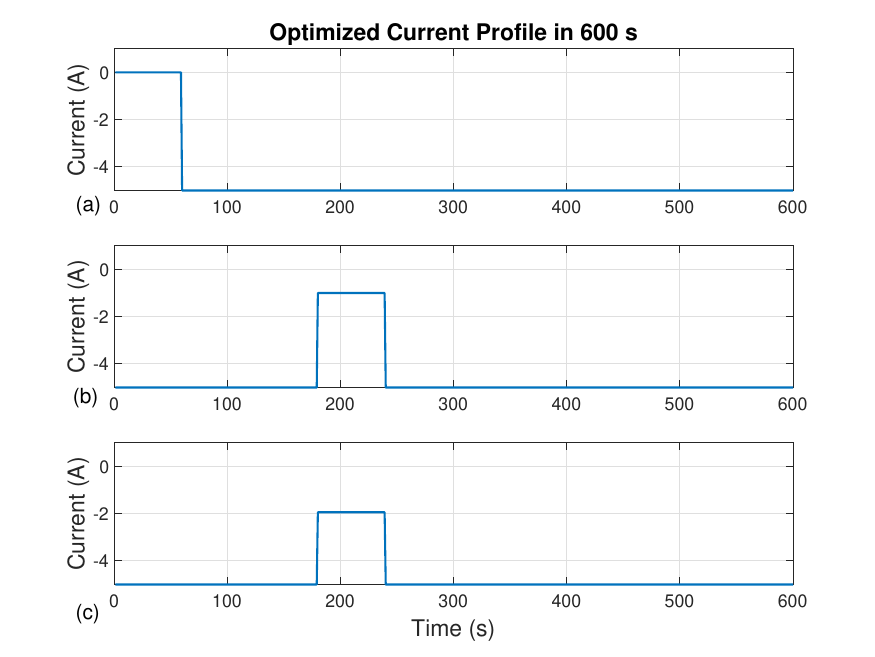}
      \caption{Optimized current excitation with a) D-optimality b) E-optimality and c) A-optimality}
      \label{fig:600s_all}
\end{figure}

Validation of the proposed reference voltage method for battery SOH estimation in simulation is presented in this section.
We used the DFN model to emulate the dynamics of a LGM 50 21700 Nickel-Manganese-Cobalt battery for data generation.
The parameters used in the DFN model were adopted from Chen et al. \cite{chen2020development}.
Additionally, the SPMe model is used for estimation in this work, which was parameterized at BOL using the same set of parameters as the DFN.
The battery was simulated at three different degradation levels in addition to the BOL condition.
These degradation levels represent varying degrees of battery aging, where the health-related parameters, including $\varepsilon_{s,p}$, $\varepsilon_{s,p}$, $\beta_{p,0\%}$, and  $\beta_{n,0\%}$, were adjusted to reflect different SOH.
The changes in these target parameters relative to their BOL values under each degradation level are summarized in Table \ref{tb:degradation_level}.
The optimized excitation profiles for D-optimality, E-optimality and A-optimality are illustrated in Figure \ref{fig:600s_all} (a), (b) and (c) respectively.
For comparison, a standard 1C CC charging profile over the same 600-second period was also applied to generate the data for estimation as comparison.

\begin{table}
\footnotesize
\centering
\caption{Summary of Degradation Levels} 
\label{tb:degradation_level}
\begin{tabular}{ c ccc ccc cccc} 
\toprule
\multirow{2}{*}{Degradation level} & \multicolumn{4}{c}{Ratio to initial value} \\\cmidrule{2-5}
   & $\varepsilon_{s,p}$ & $\varepsilon_{s,n}$ &  $\beta_{p,0\%}$ & $\beta_{n,0\%}$ \\ \midrule
5\%  & 0.95 & 0.95 & 0.975 & 0.975     \\                
10\% & 0.90 & 0.90 & 0.95 & 0.95     \\                 
20\% & 0.80 & 0.80 & 0.95 & 0.95     \\
\bottomrule                 
\end{tabular}
\end{table}

We first examine the results under the excitation optimized using D-optimality, which maximizes the determinant of the Fisher information matrix.
The joint estimation of target parameters was performed using both the proposed reference voltage trajectory method in Eqn. \eqref{eq:new opt}, and conventional method without uncertainty compensation in Eqn. \eqref{eq:conventional opt}.
The optimal parameters were found using the IPOPT solver in CasADi.
Table \ref{tb:estimation_Dopt} summarizes the estimation results. 
For every parameter and degradation level, the reference voltage method achieved significantly lower estimation errors compared to the conventional approach.
Notably, the estimation errors are around 1\% under the 600-second testing data, even at a degradation level of 20\%.
This demonstrates the robustness and accuracy of the reference voltage method in reducing estimation error across all key SOH parameters and at all degradation levels.  

As a comparison, in Table \ref{tb:estimation_Dopt}, the estimation errors of the 1C CC charging profile are significantly larger compared to the optimized excitation for both methods, highlighting the importance of excitation optimization.
\begin{table}
\footnotesize
\centering
\caption{SOH estimation results using Voltage Trajectory based on D-optimality Design and 1C CC Charging} 
\label{tb:estimation_Dopt}
\begin{tabular}{ c ccc ccc} 
\toprule
Deg. & With ref. & \multicolumn{4}{c}{Estimation error (\%)} & \multirow{2}{*}{Excitation}\\\cmidrule{3-6}
  level  & voltage? & $\varepsilon_{s,p}$ & $\varepsilon_{s,n}$ &  $\beta_{p,0\%}$ & $\beta_{n,0\%}$ & \\ \midrule
\multirow{4}{*}{5\%}  & Y & \textbf{-2.25} & \textbf{-0.27} & \textbf{0.64} & \textbf{0.44}   
                                                          & \multirow{2}{*}{D-opt.}\\
                      & N & -8.75 & 2.92 & 3.89 & -2.50   &                        \\\cmidrule{2-7}
                      & Y & 5.25 & -2.91 & \-0.51 & -7.60 & \multirow{2}{*}{1C CC} \\
                      & N & -7.27 & 2.66 & 3.71 & 0.49    &                        \\\hline
\multirow{4}{*}{10\%} & Y & \textbf{-3.83} & \textbf{-0.56} & \textbf{1.22} & \textbf{0.80}   
                                                          & \multirow{2}{*}{D-opt.}\\
                      & N & -8.64 & 2.83 & 4.23 & -2.43   &                        \\\cmidrule{2-7}
                      & Y & 5.29 & -3.01 & -0.34 & -11.70 & \multirow{2}{*}{1C CC} \\
                      & N & -6.58 & 2.72 & 3.96 & -2.74   &                        \\\hline
\multirow{4}{*}{20\%} & Y & \textbf{-1.06} & \textbf{-0.17} & \textbf{0.96} & \textbf{0.53}   
                                                          & \multirow{2}{*}{D-opt.}\\
                      & N & -7.20 & 2.90 & 4.58 & -2.12   &                        \\\cmidrule{2-7}
                      & Y & 7.24 & -2.77 & -0.64 & -13.53& \multirow{2}{*}{1C CC} \\
                      & N & -4.57 & 3.08 & 4.02 & -2.89&                        \\
                 \bottomrule                 
\end{tabular}
\end{table}


In addition to D-optimality, we also explored the excitation optimization using E-optimality and A-optimality objectives, and their results are summarized in Tables \ref{tb:estimation_Eopt}.
While all three optimization criteria significantly improved estimation accuracy compared to the conventional 1C CC charging profile, 
the results are slightly different among them due to the fundamental principles of different metrics.
Specifically, the A-optimality maximizes the trace of the Fisher information matrix, i.e. the sum of all eigenvalues,
which often puts more emphasis on the largest eigenvalues, hence giving good best-case errors (smallest error among all parameters). 
Meanwhile, the E-optimality maximizes the smallest eigenvalue of the FI matrix, essentially trying to controlling the worst-case errors (largest errors among all parameters). 
Finally, D-optimality maximizes the determinant of the FI matrix, i.e. the product of all eigenvalues, 
aiming at striking a balance among all eigenvalues (parameters). 

\begin{table}
\footnotesize
\centering
\caption{SOH estimation results using Voltage Trajectory based on E-optimality and A-optimality Design} 
\label{tb:estimation_Eopt}
\begin{tabular}{ c ccc ccc} 
\toprule
Deg. & With ref. & \multicolumn{4}{c}{Estimation error (\%)} & \multirow{2}{*}{Excitation}\\\cmidrule{3-6}
  level  & voltage? & $\varepsilon_{s,p}$ & $\varepsilon_{s,n}$ &  $\beta_{p,0\%}$ & $\beta_{n,0\%}$ & \\ \midrule
\multirow{4}{*}{5\%}  & Y &  -0.64 & -0.09 & 0.32 & -2.92 & \multirow{2}{*}{E-opt.}\\
                      & N &  0.07 & 6.97 & 1.27 & -1.44   &                        \\\cmidrule{2-7}
                      & Y &  -1.11 & -0.26 & 0.48 & -2.80 & \multirow{2}{*}{A-opt.} \\
                      & N &  -6.26 & 5.79 & 2.88 & 0.02  &                        \\\hline
\multirow{4}{*}{10\%} & Y &  -0.60 & -0.09 & 0.41 & 0.31 & \multirow{2}{*}{E-opt.}\\
                      & N &  1.90 & 7.19 & 1.01 & 1.95   &                        \\\cmidrule{2-7}
                      & Y &  -1.54 & 0.42 & 0.73 & 0.55  & \multirow{2}{*}{A-opt.} \\
                      & N &  -4.90 & 5.81 & 2.87 & 3.52  &                        \\\hline
\multirow{4}{*}{20\%} & Y &  2.41 & 0.50 & 0.004 & 0.23  & \multirow{2}{*}{E-opt.}\\
                      & N &  4.76 & 7.67 & 0.79 & 2.90   &                        \\\cmidrule{2-7}
                      & Y &  0.74 & -0.15 & 0.61 & 0.66  & \multirow{2}{*}{A-opt.} \\
                      & N &  -4.28 & 5.50 & 3.51 & 5.04  &                        \\
\bottomrule
              
\end{tabular}
\end{table}



\section{CONCLUSIONS}


In this paper, we proposed a novel reference voltage trajectory-based method for robust and accurate estimation of battery SOH.
The method leverages a reference voltage trajectory measured at the beginning of life under a specifically designed current excitation to account for modeling uncertainties that arise during parameter estimation.
By subtracting the modeling uncertainty from the measured voltage response of the degraded battery under the same excitation, we can reduce the impact of modeling uncertainties and improve the estimation accuracy of key health parameters, including the active material volume fraction ($\varepsilon_{s}$) and lithium-ion stoichiometry at 0\% SOC ($\beta_{0\%}$).
These parameters provide indication for the extent of prominent battery degradation mechanisms, including LAM and LLI. 
We further studied the design of the reference trajectory based on various optimality criteria of the parameter Fisher information.
Simulation studies demonstrated the effectiveness of the proposed method under different degradation levels, reducing estimation errors to as low as 1\% within a short testing period of 600 seconds, even under 20\% battery degradation. 
The easy-to-implement nature of the approach makes it appealing for applications such as battery repurposing and online diagnostics.
Future work could explore the application of this method to the real-world battery aging scenario with a broader range of battery chemistries and degradation paths.

\addtolength{\textheight}{-12cm}   





\section*{ACKNOWLEDGMENT}

We appreciate the funding support from the NSF CAREER Program (Grant No.2046292) and the NASA HOME Space Technology Research Institute (Grant No.80NSSC19K1052). 


\bibliographystyle{IEEEtran}
\bibliography{Reference}

\end{document}